\documentclass[a4paper,12pt]{iopart}
\input{epsf}
\usepackage{iopams}
\usepackage[dvips]{graphicx}
\begin{document}

\title[Subwavelength imaging of light by
arrays of metal-coated semiconductor nanoparticles...]
{Subwavelength imaging of light by arrays of metal-coated
semiconductor nanoparticles: a theoretical study}

\author{V.~Yannopapas}
\address{Department of Materials Science, University of Patras,
GR-26504 Patras, Greece}
\ead{vyannop@upatras.gr}
\date{\today}

\begin{abstract}
Rigourous calculations of the imaging properties of metamaterials
consisting of metal-coated semiconductor nanoparticles are
presented. In particular, it is shown that under proper choice of
geometric and materials parameters, arrays of such particles
exhibit negative refractive index within the region of the
excitonic resonance of the semiconductor. The occurrence of
negative refractive index is predicted by the extended
Maxwell-Garnett theory and confirmed by a layer-multiple
scattering method for electromagnetic waves. By using the same
method it is shown that within the negative refractive-index band,
arrays of such nanoparticles amplify the transmitted near-field
emitted while simultaneously narrow down its spatial profile
leading to subwavelength resolution. The effect of material losses
to the imaging properties of the arrays is also addressed.
\end{abstract}

\pacs{42.70.Qs, 42.25.Bs, 78.67.Pt, 78.67.Bf, 73.20.Mf}
\submitto{\JPCM} \maketitle

\section{Introduction}
\label{intro}

According to a recent definition \cite{lakhtakia_opn},
metamaterials are artificial materials which exhibit response
characteristics that are not observed in the individual responses
of its constituent materials. The most fascinating class of
metamaterials are those exhibiting simultaneously negative
permittivity $\epsilon$ and permeability $\mu$, i.e., a negative
refractive index (NRI) \cite{veselago,pendry}. The usual approach
of designing a NRI metamaterial is to combine a 'magnetic'
sublattice (one which exhibits negative $\mu$) of miniaturized RLC
circuits, e.g. split-ring resonators, with an 'electric' one
(exhibiting negative $\epsilon$) of thin metallic wires
\cite{eleftheriades}. However, an alternative route has been
recently suggested where the electric and magnetic sublattices are
occupied by units of less elaborate geometry such as cylinders or
spheres made from resonant materials (ionic, semiconducting or
plasmonic materials) displaying high-refractive index within a
specific frequency window
\cite{obrien,holloway,huang,vyam,felbacq,wheeler1,wheeler2,jylha,mackay,yv_prb,y_diso,rockstuhl}.
The magnetic activity of the cylinders/ spheres lies within the
region of Mie resonances resulting from the enhancement of the
displacement current inside each sphere which, in turn, gives rise
to a macroscopic magnetisation of the whole structure. The
electric activity is attributed to the large polarization induced
to the sphere due to the giant dielectric permittivity around the
resonance frequency of a given dielectric function, e.g.
polariton, plasmon or exciton resonance. The above theoretical
suggestions have been recently verified experimentally for arrays
of dielectric particles of millimeter \cite{peng} and micrometer
\cite{schuller} size. Such arrays can be miniaturized to the
nanometre scale in the form of nanoparticle superlattices
\cite{self-assembly}.

Perhaps, the most paramount application of NRI metamaterials is
their imaging property; as has been pioneered by Pendry
\cite{pendry} and verified experimentally, a planar slab of
metamaterial with NRI can overcome the standard diffraction limit
in imaging by focusing the far-field via negative refraction
\cite{parazzoli,houck} and by amplifying the near-field
\cite{lagarkov,grbic,aydin} by surface-plasmon excitation, a
possibility which promises the realization of a perfect lens.
Near-field amplification is also feasible separately for
S-polarized waves in the magnetostatic limit using solely RLC
circuits such as split-ring resonators and Swiss rolls
\cite{wiltshire_1,wiltshire_2} and for P-polarized waves in the
electrostatic limit using periodic structures with metallic
components such as rods \cite{ono}, wires \cite{fedorov}, and
spheres \cite{alitalo}.

In this work we study the imaging properties of three-dimensional
(3D) NRI metamaterials consisting of metal-coated semiconductor
spheres. As has been recently shown, arrays of semiconductor
spheres with strong excitonic oscillation strength (CuCl, ${\rm
Cu}_{2}{\rm O}$) can exhibit negative permeability in the visible
region, around the exciton resonance \cite{yv_prb}. In order to
realize an NRI structure in the regime of negative $\mu$, the
array of semiconductor spheres can be (a) either combined with an
array of metallic spheres (negative $\epsilon$) in a way that a
binary NRI structure is created \cite{vyam}, or (b) by coating the
semiconductor spheres with a metallic shell \cite{wheeler2,y_pss}.
The above structures have been studied by employing both
effective-medium treatments \cite{holloway,jylha,mackay} and more
rigorous electromagnetic approaches \cite{vyam,wheeler2}. By
exploiting the occurrence of NRI in the case of arrays of
metal-coated semiconductor nanoparticles, we will show that
near-field amplification along with subwavelength resolution of an
image can be achieved in such structures, for both polarization
modes. We will also elucidate the role of absorption which is
intrinsic in the materials constituting the NRI metamaterial under
study. The paper is organized as follows. In section \ref{theory}
the basic theoretical tools for studying the imaging properties of
arrays of spherical inclusions is presented. Section \ref{results}
applies the theory to the case of metal-coated semiconductor
spheres and section \ref{conclusion} concludes the paper.

\section{Theory}
\label{theory}

In this work, we are dealing with finite slabs of metamaterials
consisting of a number of planes of spheres with the same 2D
periodicity. In order to probe the imaging properties of these
structures, we consider the electric field emitted by a localized
source, namely, that radiated by a small, linear, infinitely thin,
center-fed antenna. The antenna lies along the $z$-axis from
$-d/2\leq z \leq d/2$ with current $I$. We expand the electric
field radiated by the antenna as a series of spherical waves
\cite{jackson}
\begin{equation}
{\bf E}({\bf r})=\sum_{l=1}^{\infty}\sum_{m=-l}^{l}
 \left\{a_{H l m}
h_{l}^{+} (qr){\bf X} _{l m}({\bf \hat r})+a_{E l m}\frac{{\rm
i}}{q}\nabla\times \left[h_{l}^{+}(qr){\bf X} _{l m}({\bf \hat
r})\right]\right\}\;, \label{eq:multem}
\end{equation}
${\bf X} _{l m}({\bf \hat r})$ are the so-called vector spherical
harmonics \cite{jackson} and $h_{l}^{+}$ are the spherical Hankel
functions of order $l$. $q=\omega/c$, where $c=1/\sqrt{\mu
\epsilon \mu_0 \epsilon_0}=c_{0}/\sqrt{\mu \epsilon}$ is the
velocity of light in the medium surrounding the antenna. The
magnetic multipole coefficients $a_{H l m}$ are zero whilst the
electric ones are given by \cite{naus}
\begin{equation}
a_{E l m}=\frac{I}{\pi d} \sqrt{\frac{4 \pi (2l+1)}{l(l+1)}}
(\frac{qd}{2})^2 j_{l}(\frac{qd}{2}) \label{eq:alm_antenna}
\end{equation}
provided that $l$ is odd and $m=0$. $j_{l}$ in
equation~(\ref{eq:alm_antenna}) denote the spherical Bessel
functions. Note that the above formula is valid in the case where
the antenna is much smaller than the wavelength \cite{naus}, i.e.,
$qd\ll1$.

\begin{figure}[h]
\centerline{\includegraphics[width=8cm]{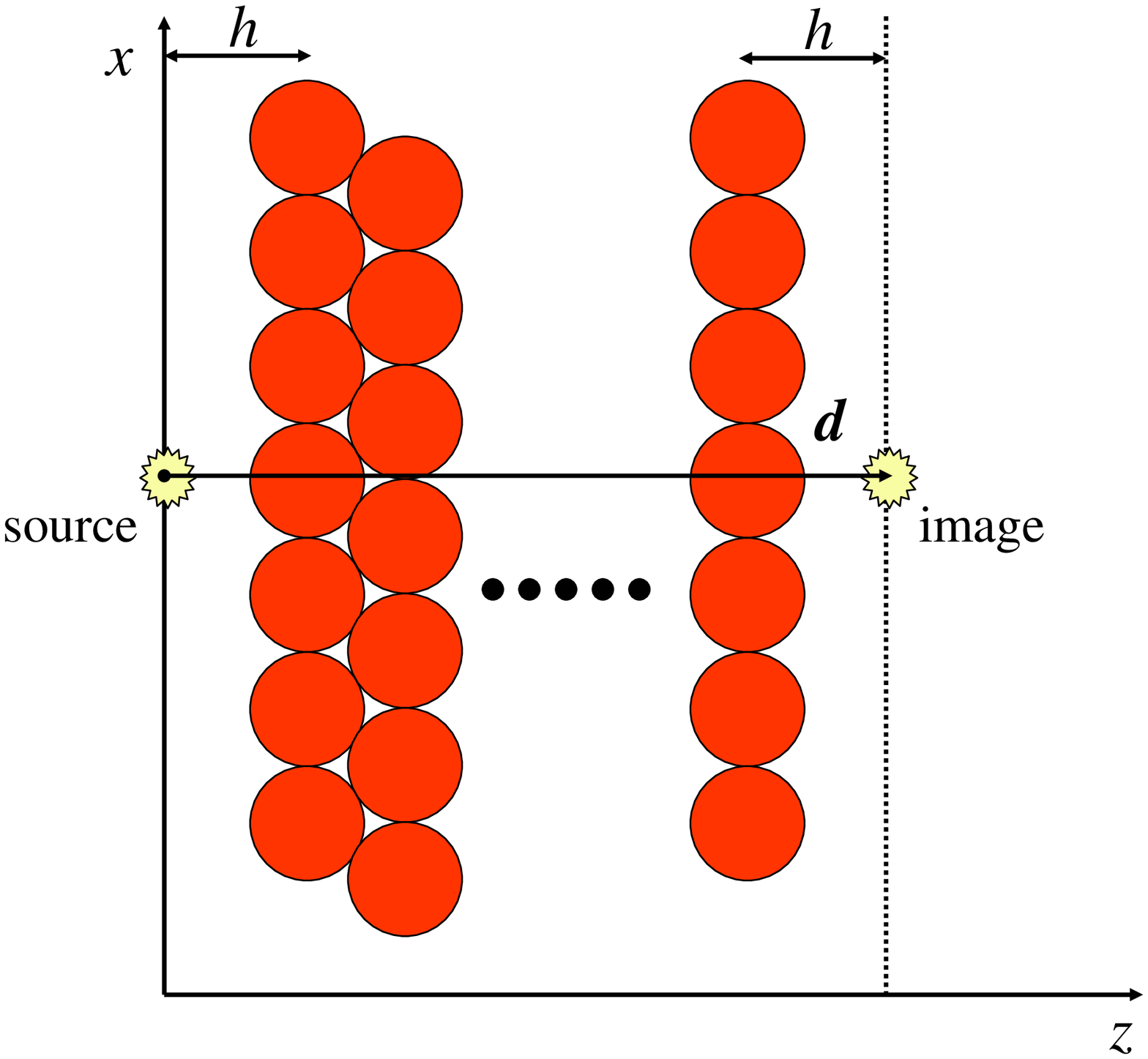}} \caption{(Color
online) Calculation setup.} \label{fig1}
\end{figure}

Since we wish to study the transmission of the above field through
a slab of a number of periodic planes of spheres, it is
advantageous to transform the field of equation~(\ref{eq:multem})
to a basis of plane waves consistent with the 2D periodicity of
the planes of spheres. If the linear antenna is placed to the left
of the slab (see the calculation setup in figure~\ref{fig1}), then
the field radiated to the right and is incident on the slab is
written as (assuming the center of coordinates be located at the
localized source)\cite{cpa3d}
\begin{equation}
{\bf E}^{inc\ +}({\bf
r})=\frac{1}{S_{0}}\int\int_{SBZ}d^{2}k_{\parallel}\sum_{{\bf
g}}{\bf E}^{inc\ +}_{{\bf g}}({\bf k}_{\parallel}) \exp({\rm
i}{\bf K}^{+}_{{\bf g}}\cdot{\bf r}) \label{outp}
\end{equation}
with
\begin{equation}
E^{inc\ +}_{{\bf g};i}({\bf k}_{\parallel})= \sum_{l=1}^{\infty}
\sum_{m=-l}^{l} \sum_{P=E,H} \Delta_{Plm;i}({\bf K}^{+}_{{\bf g}})
a_{Plm} \label{ampoutp}
\end{equation}
where $i=1,2$ are the two independent polarizations (polar and
azimuthal) which are normal to the wavevector
\cite{skm92,comphy98,comphy00}
\begin{equation}
\mathbf{K}_{\mathbf{g}}^{+}= \left(
\mathbf{k}_{\parallel}+\mathbf{g},\ \left[
q^{2}-\left(\mathbf{k}_{\parallel}+\mathbf{g}\right)^{2} \right]
^{1/2} \right). \label{eq:gk}
\end{equation}
The vectors ${\bf g}$ denote the reciprocal-lattice vectors
corresponding to the 2D periodic lattice of the plane of spheres
and ${\bf k}_{\parallel}$ is the reduced wavevector which lies
within the Surface Brillouin Zone (SBZ) associated with the
reciprocal lattice \cite{skm92,comphy98,comphy00}. When
$q^{2}<(\mathbf{k}_{\parallel}+\mathbf{g})^{2}$, the wavevector of
equation~(\ref{eq:gk}) defines an evanescent wave.
 The coefficients ${\bf
\Delta}_{Plm}$ are given by
\begin{eqnarray} {\bf
\Delta}_{El m}({\bf K}_{{\bf g}}^{+})&=&\frac{2\pi
(-\mathrm{i})^{l}}{q_{h} A_0 K_{{\bf g} z}^+\sqrt{l (l
+1)}}\biggl\{\mathrm{i}
\left[\alpha_{l}^{-m}\;e^{\mathrm{i}\phi}\;Y_{l}^{m-1}(\hat{{\bf
K}}_{{\bf g}}^{+})-\alpha_{l}^{m}\;e^{-\mathrm{
i}\phi}\;Y_{l}^{m+1}(\hat{{\bf K}}_{{\bf g}}^{+})\right]\hat{{\bf
e}}_{1} \nonumber
\\&-&\left[\alpha_{l}^{-m}\cos \theta \;e^{\mathrm{
i}\phi}\;Y_{l}^{m-1}(\hat{{\bf K}}_{{\bf g}}^{+})-m\sin \theta\;
Y_{l}^{m}(\hat{{\bf K}}_{{\bf g}}^{+}) \right. \nonumber \\
&+& \left. \alpha_{l}^{m} \cos \theta
\;e^{-\mathrm{i}\phi}\;Y_{l}^{m+1}(\hat{{\bf K}}_{{\bf
g}}^{+})\right]\hat{{\bf e}}_{2}\biggr\}\;,
\nonumber \\
 {\bf \Delta}_{H l m}({\bf K}_{{\bf g}}^{+}) &=& \frac{2\pi
(-\mathrm{i})^{l}}{q A_0 K_{{\bf g} z}^+\sqrt{l (l +1)}} \biggl\{
\left[ \alpha_{l}^{-m}\cos \theta \;e^{\mathrm{
i}\phi}\;Y_{l}^{m-1}(\hat{{\bf K}}_{{\bf g}}^{+}) \right.
\nonumber \\
&-&m \left. \sin \theta \; Y_{l}^{m}(\hat{{\bf K}}_{{\bf g}}^{+})
+ \alpha_{l}^{m} \cos \theta
\;e^{-\mathrm{i}\phi}\;Y_{l}^{m+1}(\hat{{\bf K}}_{{\bf
g}}^{+})\right] \hat{{\bf e}}_{1} \nonumber \\  &+&\mathrm{i}
\left[\alpha_{l}^{-m}\;e^{\mathrm{i}\phi}\;Y_{l}^{m-1}(\hat{{\bf
K}}_{{\bf g}}^{+})-\alpha_{l}^{m}\;e^{-\mathrm{
i}\phi}\;Y_{l}^{m+1}(\hat{{\bf K}}_{{\bf g}}^{+})\right]\hat{{\bf
e}}_{2}\biggr\}\;, \label{Dlm}
\end{eqnarray}
where $\theta$, $\phi$ denote the angular variables $(\hat{{\bf
K}}_{\mathbf{g}}^{+})$ of $\mathbf{K}_{\mathbf{g}}^{+}$ and
$A_{0}$ is the area of the unit cell of the 2D lattice occupied by
the spheres. $Y_{l}^{m}$ denotes a spherical harmonic as usual,
$\hat{\mathbf{e}}_{1}$, $\hat{\mathbf{e}}_{2}$ are the polar and
azimuthal unit vectors, respectively, which are perpendicular to
$\mathbf{K}_{\mathbf{g}}^{+}$. $\alpha_{l}^{m}$ are given by
$\alpha_{l}^{m}=\frac{1}{2}\left[(l-m)(l+m+1)\right]^{1/2}$. The
incident field of equation~(\ref{outp}) will be partly transmitted
through the slab under study. The transmitted field will be given
by
\begin{equation}
{\bf E}^{tr \ +}({\bf
r})=\frac{1}{S_{0}}\int\int_{SBZ}d^{2}k_{\parallel}\sum_{{\bf
g}}{\bf E}^{tr \ +}_{{\bf g}}({\bf k}_{\parallel}) \exp[{\rm
i}{\bf K}^{+}_{{\bf g}}\cdot({\bf r}-{\bf d})] \label{trans}
\end{equation}
with
\begin{equation}
E^{tr\ +}_{{\bf g};i}({\bf k}_{\parallel})= \sum_{{\bf g}', i'}
Q^{I}_{{\bf g} i;{\bf g}' i'} E^{inc\ +}_{{\bf g}'i'}({\bf
k}_{\parallel}) \label{amp_trans}
\end{equation}
${\bf d}$ is a vector joining the source to the image (see
figure~\ref{fig1}). The transmission matrix ${\bf Q}^{I}$
appearing in equation~(\ref{amp_trans}) is calculated within the
framework of the layer-multiple-scattering method which is an
efficient computational method for the study of the EM response of
three-dimensional photonic structures consisting of nonoverlapping
spheres \cite{skm92,comphy98,comphy00} and axisymmetric
non-spherical particles \cite{nonsph}. The
layer-multiple-scattering method is ideally suited for the
calculation of the transmission, reflection and absorption
coefficients of an electromagnetic (EM) wave incident on a
composite slab consisting of a number of layers which can be
either planes of non-overlapping particles with the same 2D
periodicity or homogeneous plates. For each plane of particles,
the method calculates the full multipole expansion of the total
multiply scattered wave field and deduces the corresponding
transmission and reflection matrices in the plane-wave basis. The
transmission and reflection matrices of the composite slab are
evaluated from those of the constituent layers. By imposing
periodic boundary conditions one can also obtain the (complex)
frequency band structure of an infinite periodic crystal. The
method applies equally well to non-absorbing systems and to
absorbing ones. Its chief advantage over the other existing
numerical methods lies in its efficient and reliable treatment of
systems containing strongly dispersive materials such as
Drude-like and polaritonic materials.

The calculation of the incident [equation~(\ref{outp})] as well as
the transmitted field [equation~(\ref{trans})], requires a
numerical integration over the entire SBZ. In the example examined
in the next section, the spheres in all planes occupy the sites of
a square lattice and, therefore, the SBZ is also a square. The SBZ
integration of equations~(\ref{outp}) and (\ref{trans}) is
performed by subdividing progressively the SBZ into smaller and
smaller squares, within which a nine-point integration formula
\cite{abramo} is very efficient. Using this formula we managed
excellent convergence with a total of 73728 points in the SBZ.
Also, the inclusion of 13 reciprocal-lattice ${\bf g}$-vectors
along with an angular-momentum cutoff $l_{max}=4$ provided
converged results.

\section{Results}
\label{results}

We consider a 3D array of closed-packed CuCl nanoparticles of
radius $S=28$~nm; CuCl exhibits a ${\rm Z_{3}}$ exciton line at
386.93~nm \cite{artoni}. Around the exciton frequencies, the
dielectric function of the above semiconductors is given by
\begin{equation}
\epsilon_{s} (\omega) = \epsilon_{\infty} + A
\gamma/(\omega_{0}-\omega-{\rm i} \gamma). \label{eq:CuCl_diel}
\end{equation}

The constant $A$ is proportional to the exciton oscillator
strength and for CuCl, $A=632$. The rest of the parameters for
CuCl are \cite{artoni}: $\epsilon_{\infty}=5.59$, $\hbar
\omega_{0}=3.363$~eV, and $\hbar \gamma=5 \cdot 10^{-5}$~eV. The
small value of the loss factor $\gamma$ implies a very narrow
exciton linewidth. The magnetic permeability $\mu_{s}$ of CuCl is
unity. Arrays of CuCl NPs can be fabricated by colloidal
crystallization \cite{orel} and ion implantation techniques
\cite{fukumi}. The CuCl NPs are coated with a metal of nanometre
thickness, $\ell=0.10S=2.8$~nm (nanoshell). Such hybrid
metal-semiconductor nanoparticles have already been synthesized in
the laboratory (for a recent review see Ref.~\cite{cozzoli}). The
dielectric function of the metal is assumed to be described by the
Drude model, i.e.,
\begin{equation}
\epsilon_{m}=1-\frac{\omega_{p}^{2}}{\omega (\omega +{\rm i}
\gamma)}. \label{eq:drude}
\end{equation}
In order to achieve NRI, $\omega_{p}$ is taken to be
$\omega_{p}=1.05 \omega_{0}$. For the loss factor $\gamma$ we have
taken a typical value of $\gamma/ \omega_{p}=0.01$ \cite{vyam}.

\begin{figure}[h]
\centerline{\includegraphics[width=8cm]{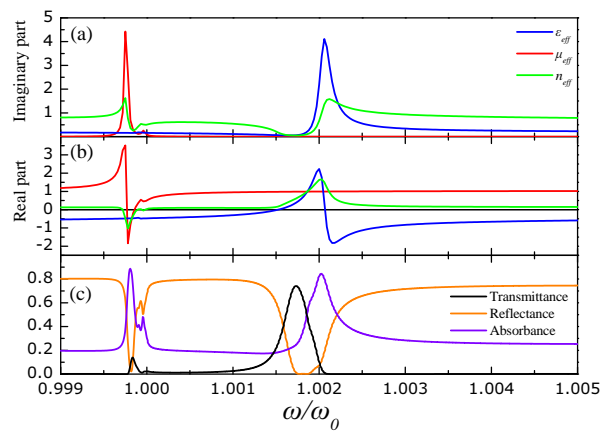}} \caption{(Color
online) The imaginary (a) and real (b) parts of the effective
parameters $n_{eff}$ (solid line), $\epsilon_{eff}$ (dashed line)
and $\mu_{eff}$ (dotted line) for a 3D array of close-packed
metal-coated CuCl spheres of radius $S=28$~nm and coating
thickness $\ell=2.8$~nm. (b) Transmittance, reflectance and
absorbance of light incident normally on a slab consisting of 4
(001) fcc planes of the above nanospheres, as calculated by the
layer-multiple scattering method \cite{skm92,comphy98,comphy00}. }
\label{fig2}
\end{figure}

The effective permittivity $\epsilon_{eff}$ and permeability
$\mu_{eff}$ of such a structure can be calculated by the extended
Maxwell-Garnett theory \cite{doyle,ruppin} which encompasses
elements of the Mie theory within the formulae of the
$\epsilon_{eff}$, $\mu_{eff}$. As such, the extended
Maxwell-Garnett theory agrees very well with more rigorous
approaches \cite{vyam,wheeler1, wheeler2,yv_prb,y_diso}.
Figure~\ref{fig2}a shows the real and imaginary parts of
$\epsilon_{eff}$, $\mu_{eff}$, and $n_{eff}$ for the above
described system of metal-coated CuCl nanoparticles. One clearly
observes a region of negative $\Re \epsilon_{eff}$ for $\omega/
\omega_{0} < 1.0015$. One can also identify a narrow frequency
region around $\omega/ \omega_{0} \sim 0.9998$, where $\Re
\mu_{eff}<0$ which is entirely within the region of negative $\Re
\epsilon_{eff}$. Within this region, $\Re n_{eff}<0$ and a NRI
band occurs. Note that for the calculation of the $n_{eff}$ the
imaginary parts of $\epsilon_{eff}$, $\mu_{eff}$ have been also
taken into account by choosing the value of
$n_{eff}=\sqrt{\epsilon_{eff} \mu_{eff}}$ which possesses positive
imaginary part (the structures under study are made from passive
materials).

In order to verify the validity of the effective medium parameters
depicted in figure~\ref{fig2}a,b we have also employed the
layer-multiple scattering method \cite{skm92,comphy98,comphy00}
(briefly outlined in section \ref{theory}) in order to calculate
the transmittance of light incident normally on a finite slab of
an fcc crystal of the above nanospheres. We have chosen the fcc
lattice since this type of lattice can satisfy a close-packed
arrangement of spheres. The slab consists of 4 (001) fcc planes of
nanospheres and the respective transmittance is shown in
figure~\ref{fig2}c. It is evident that within the NRI of
figure~\ref{fig2}b, the transmittance shows a maximum which is
what one expects from a slab of NRI \cite{pendry}. As light
propagates through the slab, it is also attenuated due to the
intrinsic losses of the constituent materials (CuCl and
Drude-metal) as it is evident from the corresponding maximum of
the absorbance curve of figure~\ref{fig2}c within the NRI band. It
is also evident that there is also a wider transmittance peak from
$\omega/ \omega_{0}\simeq 1.0015$ to 1.002 which, however,
corresponds to a region of refractive index with positive real
part (the matching of these two frequency regions is not perfect
as figure~\ref{fig2}a is based on an effective medium
approximation whilst figure~\ref{fig2}b to an exact theory). In
between these peaks as well as above the second peak, the
transmittance is suppressed as only one of $\Re \epsilon_{eff}$
and $\Re \mu_{eff}$ becomes negative giving rise to a practically
imaginary refractive index (see figures~\ref{fig2}a,b).

\begin{figure}[h]
\centerline{\includegraphics[width=8cm]{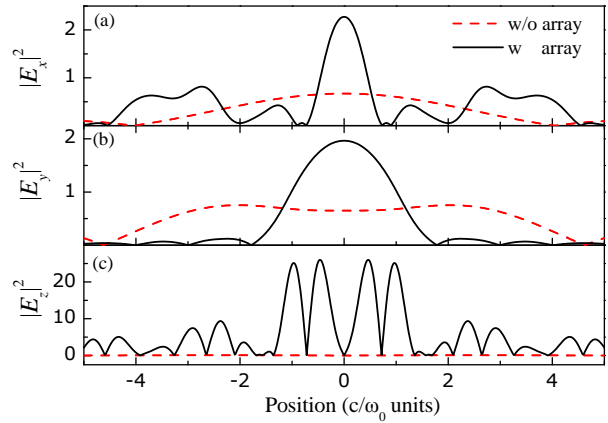}} \caption{(Color
online) Distribution of $|E_{x}|^{2}$ (a), $|E_{y}|^{2}$ (b), and
$|E_{z}|^{2}$ (c) (in arbitrary units) along the line $y=0$ in the
image plane (see figure~\ref{fig1}), for frequency $\omega/
\omega_{0}=0.99984$. The solid lines refer to the field
distribution from a localized source (antenna) placed at a
distance $h=0.6 c/ \omega_{0}$ from the center of a single plane
of close-packed metal-coated CuCl nanospheres (see
figure~\ref{fig1}). The dashed lines refer to the free-space case
(no spheres present).} \label{fig3}
\end{figure}

In order to probe the imaging properties of the above systems, we
have applied the formalism developed in section \ref{theory} for
the case of a single plane of metal-coated CuCl nanospheres. As a
localized source of light we have considered a linear, infinitely
thin, center-fed antenna of width $d=0.001 c/ \omega_{0}$, lying
along the $z$-axis. It is placed at a distance $h=0.6 c/
\omega_{0}$ from the center of a single plane of spheres (see
figure~\ref{fig1}). Figure~\ref{fig3} shows the distribution of
all three electric-field components along the line $y=0$ in the
image plane (see figure~\ref{fig1}), for frequency $\omega/
\omega_{0}=0.99984$ which corresponds to the low-frequency maximum
of the transmittance curve of figure~\ref{fig2}b.  For comparison,
we also show the corresponding curves in the absence of the plane
of spheres. We note that the field distributions of all figures
that follow contain both the far-field
[$q^{2}>(\mathbf{k}_{\parallel}+\mathbf{g})^{2}$ in
equation~(\ref{eq:gk})] and near-field
[$q^{2}<(\mathbf{k}_{\parallel}+\mathbf{g})^{2}$ in
equation~(\ref{eq:gk})] components. Since the chosen frequency
($\omega \omega_{0}=0.99984$) lies within the NRI band of the
array of spheres (see above) we expect that the near-field
components which are transmitted through the plane will be
amplified. And this is indeed what we observe in
figure~\ref{fig3}. We observe that {\it all} three components
components of the electric field are amplified with respect to the
case where the plane of spheres is absent. This can occur only in
the case of NRI where, as predicted by Pendry \cite{pendry}, both
polarization modes are amplified when transmitted through a planar
NRI slab. We remember that when only one of $\epsilon_{eff}<0$ or
$\mu_{eff}<0$ occurs, the near-field amplification affects one of
the polarization modes
\cite{wiltshire_1,wiltshire_2,ono,fedorov,alitalo}. In addition to
the near-field amplification of the electric field, we observe
that the FWHM for the $E_{x}-$ ($\sim \lambda_{0}/7$) and
$E_{y}-$components ($\sim \lambda_{0}/3$) is considerably smaller
than the free-space case suggesting a subwavelength imaging
operation of the structure (see below). The $E_{z}$-component
(figure~\ref{fig3}c) is much more amplified than the other two;
however, the field distribution looks much more complicated,
possibly, due to interference effects (the far-field is also
included in the calculation) and the peculiarity of the source
(the $E_{z}$-component emitted by a center-fed antenna possesses a
nodal line along the direction of the antenna axis).

\begin{figure}[h]
\centerline{\includegraphics[width=8cm]{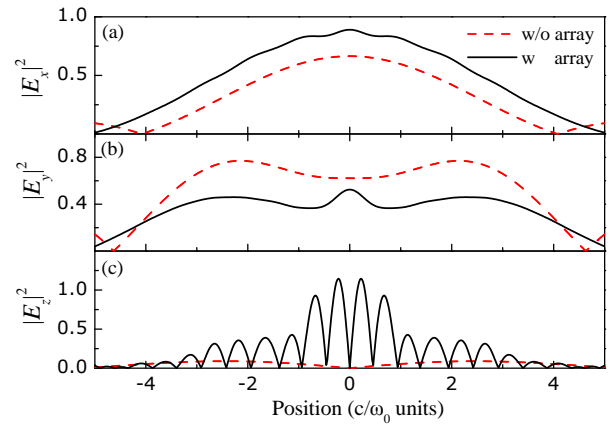}} \caption{(Color
online) Same same as in figure~\ref{fig3} but for frequency
$\omega/ \omega_{0}=1.00168$.} \label{fig4}
\end{figure}

In order to confirm the fact that the near-field amplification
shown in figure~\ref{fig3} is attributed to the occurrence of NRI,
in figure~\ref{fig4} we show the same quantities, for the same
calculation setup as in figure~\ref{fig3}, but for frequency
$\omega/ \omega_{0}=1.00168$ corresponding to the high-frequency
transmission peak of figure~\ref{fig2}b which lies out of the NRI
band. We observe that only the $E_{z}$-component
(figure~\ref{fig4}c) shows a similar trend as that of
figure~\ref{fig3}c. The $E_{y}$-component is suppressed compared
to the free-space case whilst the $E_{x}$-component is barely
amplified from the plane of nanospheres. Also, the subwavelength
profile of the field distributions is lost as the maxima are
almost as wide as those of the free-space case. So, judging from
figure~\ref{fig4}, we can infer that the field amplification for
all polarizations and the subwavelength extension of the electric
field of figure~\ref{fig3} are in accordance with the occurrence
of NRI.

\begin{figure}[h]
\centerline{\includegraphics[width=8cm]{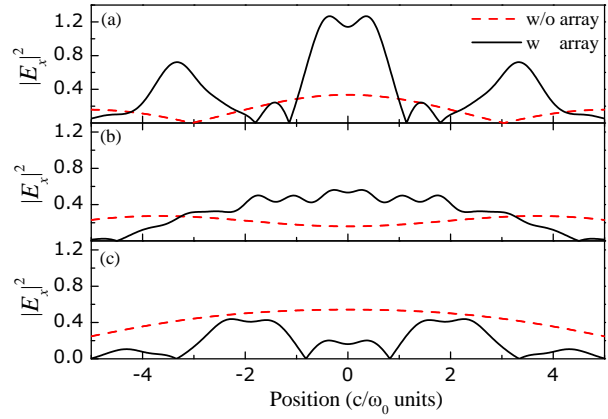}} \caption{
(Color online) Solid lines: distribution of $|E_{x}|^{2}$ (in
arbitrary units) along the line $y=0$ in the image plane (see
figure~\ref{fig1}), for frequency $\omega/ \omega_{0}=0.99984$, in
the presence of a slab consisting of 2 (a), 3 (b) and 4 (c) planes
of close-packed metal-coated CuCl nanospheres. The distance of the
source from the center of left plane of the slab is $h=0.6 c/
\omega_{0}$. The dashed lines refer to the free-space case (no
spheres present).} \label{fig5}
\end{figure}

Next, in figure~\ref{fig5}, we probe the imaging properties of
thicker slabs, i.e., slabs containing more than one planes of
nanospheres. We have chosen to depict the $E_{x}$-component but
similar effects are observed for the other two components. For the
case of two planes of nanospheres (figure~\ref{fig5}a), the field
amplification is still evident although somewhat distorted at the
center. There are also two satellite peaks evident which may
influence a potential lensing application of the structure. For a
3-planes-thick slab (figure~\ref{fig5}b) the amplification of the
transmitted field is not as dramatic as for thinner slabs
(Figs.~\ref{fig3}a and \ref{fig5}a) whilst for the 4-planes-thick
slab the amplification is evidently lost. At first glance, this
seems to be an unexpected result: the extended Maxwell-Garnett
theory \cite{doyle,ruppin} which was employed for the calculation
of $n_{eff}$ in figure~\ref{fig2}a refers to 3D collections of
scatterers and it is is supposed to be more applicable to thick
enough slabs. However, as both constituent materials of the
nanospheres (metal and semiconductor) suffer from losses
(especially the metal) we expect that thicker slabs mean longer
optical paths for the transmitted field, and, therefore, higher
losses which counterbalance the near-field amplification effect.
This is also manifested in figure~\ref{fig2}c where almost 90\% of
the incident power is absorbed within the slab (of 4 planes of
spheres). Similar effects have been also measured experimentally
for NRI metamaterials in the microwave regime \cite{aydin}.

\begin{figure}[h]
\centerline{\includegraphics[width=8cm]{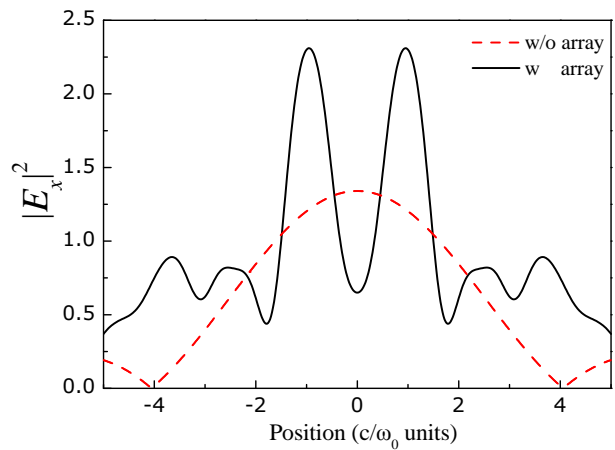}} \caption{(Color
online) Field distribution for two sources separated by distance
$c/ \omega_{0}=\lambda_{0}/ \pi$ and arranged symmetrically with
respect to the center of a sphere of a single plane of
close-packed metal-coated CuCl nanospheres (solid lines). The
sources are placed at a distance $h=0.6 c/ \omega_{0}$ from the
center of a single sphere of the plane. The dashed lines refer to
the free-space case (no spheres present).} \label{fig6}
\end{figure}

Finally, in order to study the image-resolution properties of the
array of nanospheres, in figure~\ref{fig6}, we have considered the
case of two localized sources (small center-fed antennas)
separated by a vertical subwavelength distance of $c/ \omega_{0}=
\lambda_{0}/ \pi$. The array consists of a single plane of
nanospheres since absorption is expected to deteriorate the image
resolution for thicker slabs. The sources are placed symmetrically
with respect to the center of a sphere, at a distance $h=0.6 c/
\omega_{0}$ from the plane. As it is evident from the dashed curve
of figure~\ref{fig6}, the distance between the sources is too
small to be resolved in free-space (a single peak appears).
However, in the presence of the plane of nanospheres, the
occurrence of NRI is accompanied by a dramatic increase in the
image resolution which allows for distinguishing the two sources
in space by use of visible light. We note that the image
resolution can be further improved by using smaller nanospheres,
down to radius of 15~nm. However, for CuCl spheres smaller than
that, $\mu_{eff}$ is no longer negative \cite{y_apa} around the
exciton resonance and a NRI band is not expected to occur.

\section{Conclusion}
\label{conclusion}

It has been shown that arrays of metal-coated semiconductor
nanospheres can exhibit NRI around the exciton resonance (if such
exists) of the semiconductor. The existence of NRI has been
studied by use of the extended Maxwell-Garnett theory and
confirmed by rigorous EM calculations of light transmittance.
Within the NRI band, an array of such spheres can amplify the
near-field for both polarization modes, in accordance with
Pendry's theory \cite{pendry}. The near-field amplification is
accompanied by subwavelength image resolution that is only limited
by the inherent losses of the metal and the semiconductor. The
systems studied in this work are strong candidates for NRI
metamaterials in the optical regime due to their fabrication with
existing methods of colloidal chemistry
\cite{orel,fukumi,cozzoli,gaponenko}.

\small
\normalsize

\small
\normalsize

\small
\normalsize

\small
\normalsize

\small
\normalsize

\small
\normalsize

\end{document}